Lübeck, 22.10.2017

# Genaue modellbasierte Identifikation von gynäkologischen Katheterpfaden für die MRT-bildgestützte Brachytherapie

Andre Mastmeyer

Institut für Medizinische Informatik, Universität zu Lübeck

## 1. Einleitung

Die Sterblichkeit bei gynäkologischen Krebsarten einschließlich Zervix-, Eierstock-, Vaginal- und Vulvakrebs beträgt international mehr als 6% [1]. In vielen Ländern wird die externe Strahlentherapie standardmäßig durch Brachytherapie mit hohen lokal applizierten Dosen ergänzt. Die überlegene Fähigkeit der Magnetresonanztomographie (MRT), Weichgewebe differenziert darzustellen, führt zu einem zunehmenden Einsatz dieser Bildgebungstechnik bei der intraoperativen Planung und Durchführung von Brachytherapien. Eine technische Herausforderung verbunden mit der Verwendung von MRT-Bildgebung für die Brachytherapie - im Gegensatz zur Computertomographie (CT)-Bildgebung - ist die dunkel-diffuse Darstellung und dadurch erschwerte Identifikation der Katheterpfade in den resultierenden Bildern. Dieses Problem wird durch die hier beschriebene präzise Methode zur Rückverfolgung der Katheter von der Katheterspitze ausgehend adressiert. Die durchschnittliche Identifikationszeit für einen einzelnen Katheterpfad betrug drei Sekunden auf einem Standard-PC. Die Segmentierungszeit, Genauigkeit und Präzision sind vielversprechende Indikatoren für den Wert dieser Methode für die klinische Anwendung der bildgestützten gynäkologischen Brachytherapie.

Nach einer Operation (OP) wird das gesunde Umgebungsgewebe des Tumors meist bestrahlt. Damit verringert man das Risiko, dass davon Restzellen zurückbleiben, die eine Wiederkehr des Krebsgeschwürs oder die Bildung von Metastasen - Tochtergeschwülste an anderer Stelle des Körpers - wahrscheinlich machen würden. Bei einem Tumor an Gebärmutterhals oder Prostata erfolgt die OP minimalinvasiv, das heißt, die Entfernung des Krebsgeschwürs und die Bestrahlung werden Kosten- und Risiko-vermeidend per Schlüsselloch-OP statt einer offenen OP durchgeführt.

## 2. Methode

Die hierzu verwendete Brachytherapie ist eine Form der Strahlentherapie, bei der eine umschlossene radioaktive Strahlenquelle innerhalb oder in unmittelbarer Nähe des zu bestrahlenden Gebietes im Körper platziert wird. Im Rahmen des von der Universität zu Lübeck, Institut für Medizinische Informatik, in Kooperation mit der Harvard Medical School durchgeführten Publikationsprojektes wurden Methoden für die hochgenaue und präzise Segmentierung von Katheterpfaden entwickelt. Sie ermöglichen es, während der Operation im MRT-Bild die den Tumor treffenden Pfade den Öffnungen einer Syed-Neblett-Schablone zuzuordnen (Abb. 1a). In der OP betrachtet der interventionelle Radiologie die multi-planaren Reformationen der MRT-Bilder der Katheterpfade und klickt auf das superiore Ende eines Pfades, welcher z.B. optimal in der Läsion liegt. Der Algorithmus liefert die Koordinate des passenden Schablonenloches. Dies ist für die fehlerlose Dosierung der durch die Schablone und die anliegenden Pfade eingeführten Strahlenquellen von zentraler Bedeutung (Abb. 1b). Die Deponierung der Strahler geschieht intraoperativ, bildgestützt, den Tumor sehend (Abb. 2a) und gegebenenfalls unter mehrfacher Nutzung der angelegten Katheterpfade.

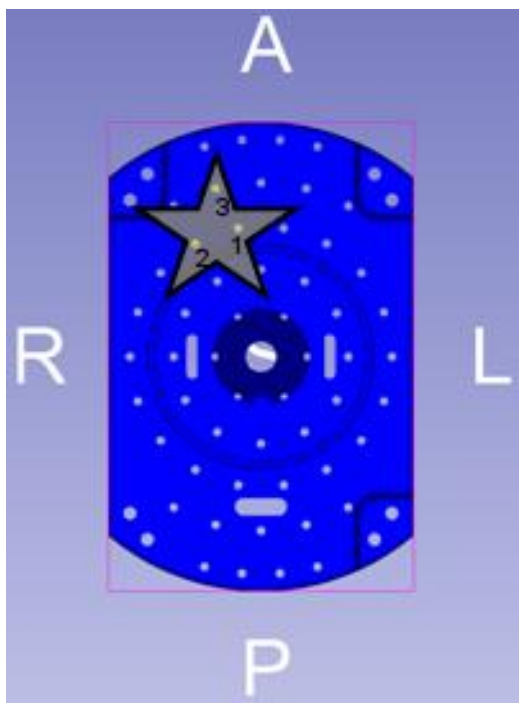

Abb. 1a: Gyn-Syed-Neblett-Schablone mit Katheter-IDs 1, 2 und 3. Katheter werden eingeführt.

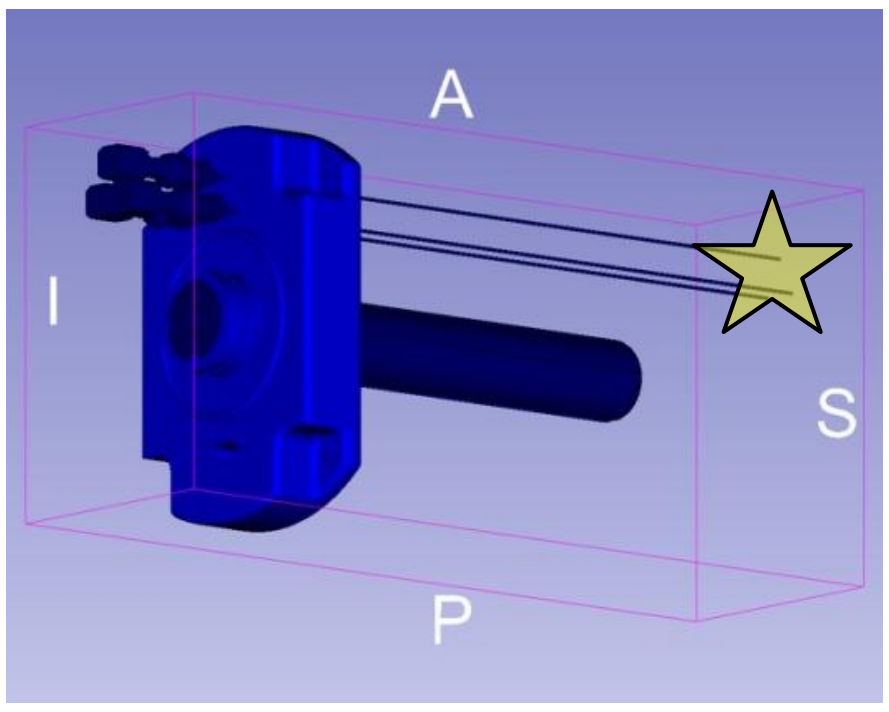

Abb. 1b: Syed-Neblett-Schablone mit drei Kathetern und Tumorläsion (gelb): Die Dosisapplikation soll intern im Patienten optimal im Tumor deponiert werden.

Im Lübecker Institut für Medizinische Informatik werden Assistenzsysteme für die intraoperative Planung und Navigation von Katheterpunktionen zur Cervix- und Prostata-Brachytherapie entwickelt, bei denen spezielle die Bildgebung unterstützenden Methoden der Nachverarbeitung (Postprocessing) zum Einsatz kommen. Die Segmentierungsalgorithmen basieren auf (1) speziellen lokalen Bildfiltern für die dunkel-diffusen Katheterartefakte, (2) auf einem mechanischen Kathetermodell [2, 3] wie es auch in aktuellen Virtual Reality Simulatoren für Nadelpunktionen [4, 5] mit virtuellen Patienenkörpermodellen [6-9] verwendet wird. In den Simulationen wird im Gegensatz zu dieser Arbeit auch die durch das Instrument verursachte Deformation des Weichteilgewebes berücksichtigt [10-12]. Hier wird mit einem Nadelmodell bereits erfolgreich der wahrscheinliche Katheterpfad mithilfe der wirkenden Katheterbiegungskräfte zur Schablone zurückverfolgt, und (3) logischem Ausschließen nicht plausibler, gruppenweiser Katheterpfadkonfigurationen (Überschneidungen).

Die Katheter werden dazu als Anreihung von kurzen starren Stäben mechanisch modelliert, die miteinander durch Torsionsfedern verbunden sind. Von der Spitze des Katheters zu seiner Basis hin sind die Katheter Kräften ausgesetzt, die in dem Modell berücksichtigt werden. Die Kathetersegmentierung in einem MRT-Bild wird durch den Benutzer initiiert, der den Ort einer Katheterspitze per Mausklick vorgibt. Diese Initialisierung ist gefolgt von einer per Modellvorgaben eingeschränkten bildeigenschaftenbezogenen Suche nach jedem dieser kurzen, dünnen, dunklen Stababschnitte. Eine Biegung zwischen benachbarten Stababschnitten von der Spitze zur Basis wird dabei immer unwahrscheinlicher. Dieser Zusammenhang wird bei der Suche durch das mechanische Modell berücksichtigt. Nach der individuellen Segmentierung aller Katheter werden diejenigen, die sich gegenseitig schneiden, automatisch identifiziert und in einem Nachbearbeitungs- und Verfeinerungsschritt korrigiert.

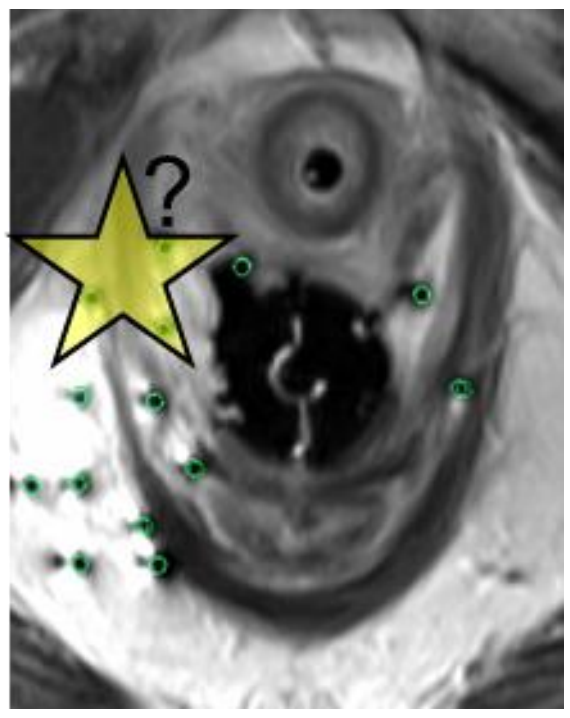

Abb. 2a: Bildkontrolle: Intraoperative axiale Schicht: Welcher Katheter gehört zu welcher ID ?

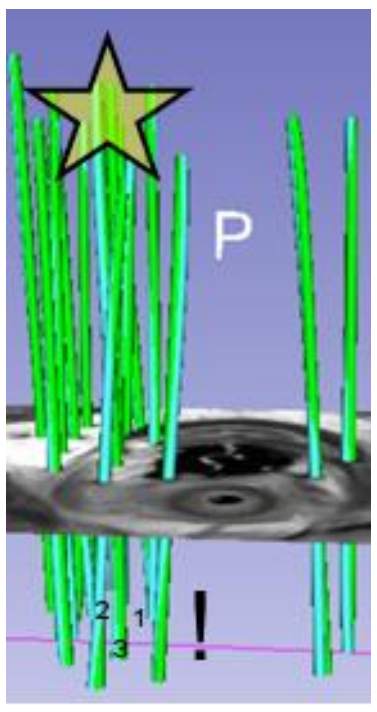

Abb. 2b: SI-Rückverfolgung ergibt IDs. Strahler korrekt lokal anwenden !

Das Hauptproblem bisher: Viele der MRT-Bilddaten enthalten nur dunkle, diffuse und dadurch sehr schwer zu deutende Katheternadeldarstellungen - also kleine, die Katheter ausmachende Bildspuren. Diese erschweren eine zuverlässige Platzierung und Dosierung der Strahlen, da der Operateur von inferior nur die Schablone für die Katheternadelneinführung und nicht das Innere des Patienten sieht. Die Artefakte befinden sich in den Bildern vom Inneren des Patienten in einem Grauwertespektrum, das im Gegensatz zur leicht erkennbaren, hellen Darstellungen in oft verwendeten CT-Aufnahmen mit zusätzlicher Dosisbelastung zu Verwechslungen mit anderen Bildstrukturen führen kann. Hier greift das unterstützende Modell ein und steuert die Segmentierung bzw. Identifikation präzise.

## 3. Ergebnisse

Die Arbeit zu dieser Publikation wurde federführend an der Universität zu Lübeck von Mai 2015 – April 2017 durchgeführt und in der renommiertesten Fachzeitschrift auf dem Gebiet der Medizinischen Bildanalyse 2017 publiziert [13]. Im Projektkontext stellt die bei Medical Image Analysis international veröffentlichte Arbeit (Impact-Faktor 5,012) der Allgemeinheit unter Berücksichtigung der besonders schwierigen dunkel-diffusen Darstellung der Katheterartefakte in MRT-Bilddaten (Abb. 2, 3) nunmehr zuverlässige Segmentierungs- und Identifikationsansätze zur Verfügung.

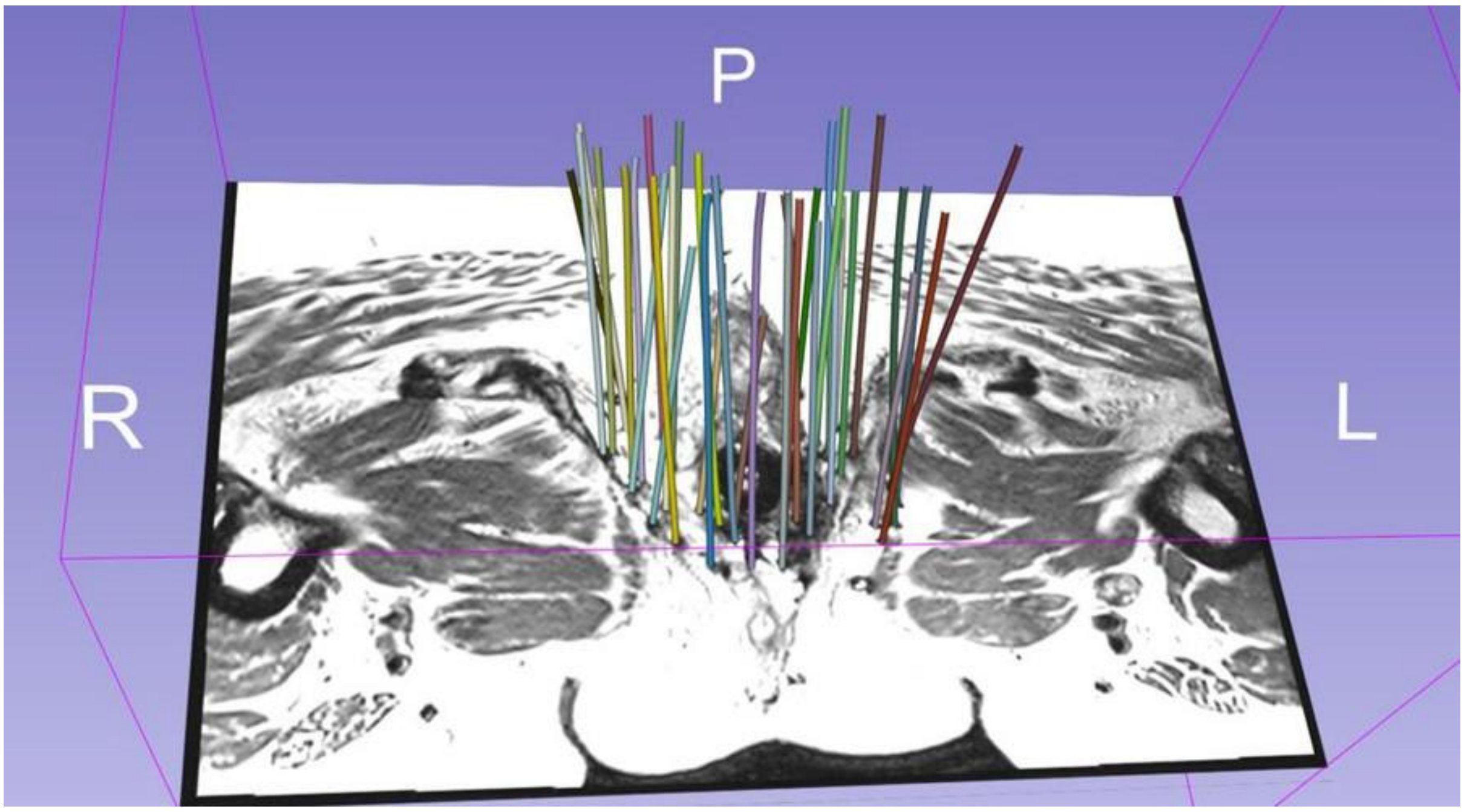

Abb. 3: Typische Kathetersituation mit zahlreichen in die Patientin eingeführten Kathetern, die mit der beschriebenen modellbasierten Methode bildnachverarbeitend identifiziert wurden.

Die umfangreiche Evaluation wurde auf 762, vorher segmentierten Referenztrajektorien (Abb. 2a, b, türkis-farbene Pfade) aus einer umfangreichen Population mit 54 intraoperativen 3D-MRT-Patientendatensätzen ausgewertet. Die erreichte Fehlergüte setzt aktuell den Goldstandard auf dem Arbeitsgebiet der bildgestützten Therapie. Es wurde die quantitative Bewertung der Genauigkeit und Robustheit vorgenommen: Im Vergleich zu expertensegmentierten Goldstandard-Kathetern ergeben sich bei 93 Prozent möglicher Identifikationsrate eine niedriger Maximalabstand 1,49 Millimeter (Hausdorff-Metrik) und exzellente Präzisionsfehler von 0,29 Millimeter (Abb. 2a, b, grüne Pfade).

## 4. Fazit

Bei Eingriffen mit Kathetern sind Chirurgen in der OP auf präzise bildgestützte, computerbasierte Assistenzsysteme angewiesen. Ein neues Verfahren verbessert nun die genaue Deutung von unklaren MRT-Bildern und macht so die minimalinvasive Krebstherapie in Zukunft sicherer und kostensparender. Die Software ist im Rahmen eines Softwarepakets Slicer als Erweiterungsmodul IGT: NeedleFinder verfügbar und wird ständig weiterentwickelt. Der Arbeitsablauf mit der Software unter der Plattform 3DSlicer ist innerhalb eines halben Tages anhand erlernbar. Zukünftig könnte bspw. noch die Modellierung der Gewebekräfte in der Katheterumgebung durch den Vorverarbeitungsschritt einer gewebeklassifizierenden Modellierung des Patienten (Segmentierung), automatisierte Pfadplanungsschritte [14, 15] und der Berücksichtigung von Patientenbewegungen [5, 16, 17] erfolgen, um das Verfahren weiter zu verbessern.

Lübeck, 22.10.2017

## Literatur

1. Ferlay, J., et al., Cancer incidence and mortality worldwide: sources, methods and major patterns in GLOBOCAN 2012. International journal of cancer, 2015. 136(5).
2. DiMaio, S.P. and S.E. Salcudean, Needle insertion modeling and simulation. IEEE Transactions on Robotics and Automation, 2003. 19(5): p. 864-875.
3. DiMaio, S.P. and S.E. Salcudean, Needle Steering and Motion Planning in Soft Tissues. IEEE Journal of Biomedical Engineering, 2005. 52(6): p. 965-974.
4. Fortmeier, D., et al., A Virtual Reality System for PTCD Simulation Using Direct Visuo-Haptic Rendering of Partially Segmented Image Data. IEEE Journal of Biomedical and Health Informatics, 2016. 20(1): p. 355-366.
5. Fortmeier, D., et al., Direct Visuo-Haptic 4D Volume Rendering Using Respiratory Motion Models. IEEE Transactions on Haptics, 2015. 8(4): p. 371-383.
6. Mastmeyer, A., D. Fortmeier, and H. Handels, Efficient Patient Modeling for Visuo-Haptic VR Simulation using a Generic Patient Atlas. Computer Methods and Programs in Biomedicine, 2016. 132: p. 161-175.
7. Mastmeyer, A., D. Fortmeier, and H. Handels, Anisotropic Diffusion for Direct Haptic Volume Rendering in Lumbar Puncture Simulation, in Bildverarbeitung in der Medizin 2012, Informatik aktuell, T. Tolxdorff, et al., Editors. 2012, Springer, Berlin Heidelberg: Berlin. p. 286 - 291.
8. Mastmeyer, A., D. Fortmeier, and H. Handels. Random Forest Classification of Large Volume Structures for Visuo-Haptic Rendering in CT Images. in Proc. SPIE Medical Imaging: Image Processing. 2016.
9. Mastmeyer, A., et al. Patch-based Label Fusion using Local Confidence-measures and Weak Segmentations. in Proc. SPIE Medical Imaging: Image Processing. 2013. Orlando, USA.
10. Fortmeier, D., A. Mastmeyer, and H. Handels, Image-based Soft Tissue Deformation Algorithms for Real-time Simulation of Liver Puncture. Current Medical Imaging Reviews, 2013. 9(2): p. 154-165.
11. Fortmeier, D., A. Mastmeyer, and H. Handels. Optimized Image-Based Soft Tissue Deformation Algorithms for Visualization of Haptic Needle Insertion. in Studies in Health Technology and Informatics: Medicine Meets Virtual Reality 20 - MMVR 2013, San Diego, California, USA, February 20-23, 2013. 2013.
12. Fortmeier, D., A. Mastmeyer, and H. Handels, GPU-Based Visualization of Deformable Volumetric Soft-Tissue for Real-Time Simulation of Haptic Needle Insertion, in Bildverarbeitung für die Medizin 2012, Informatik aktuell, T. Tolxdorff, et al., Editors. 2012, Springer, Berlin Heidelberg: Berlin. p. 117 - 122.
13. Mastmeyer, A., et al., Accurate model-based segmentation of gynecologic brachytherapy catheter collections in MRI-images. Medical Image Analysis, 2017. 42: p. 173 - 188.
14. Mastmeyer, A., et al., Ray-casting Based Evaluation Framework for Haptic Force-Feedback During Percutaneous Transhepatic Catheter Drainage Punctures. International Journal of Computer Assisted Radiology and Surgery, 2014. 9: p. 421-431.
15. Mastmeyer, A., D. Fortmeier, and H. Handels, Evaluation of Direct Haptic 4D Volume Rendering of Partially Segmented Data for Liver Puncture Simulation. Nature Scientific Reports, 2017. 7: p. 671, 1-15.
16. Mastmeyer, A., M. Wilms, and H. Handels, Interpatient Respiratory Motion Model Transfer for Virtual Reality Simulations of Liver Punctures. Journal of World Society of Computer Graphics - WSCG, 2017. 25(1): p. 1-10.
17. Mastmeyer, A., M. Wilms, and H. Handels, Interpatientenübertragung von Atemmodellen für das Virtual-Reality-Training von Punktionseingriffen, in Bildverarbeitung für die Medizin 2017, K.H. Maier-Hein, et al., Editors. 2017, Springer Vieweg, Berlin Heidelberg: Heidelberg. p. 340-345.